\documentclass[runningheads]{llncs}
\usepackage[paperheight=235mm, paperwidth=155mm,textwidth=12.2cm,textheight=19.3cm]{geometry}
\usepackage{bbding}
\usepackage{graphicx}
\usepackage{amsmath} 
\usepackage{amsfonts} 
\usepackage{amssymb}
\usepackage{stmaryrd}
\usepackage{ifthen}
\usepackage{cite}
\usepackage{todonotes} 
\usepackage{thmtools}
\usepackage{thm-restate} 
\usepackage{bm}
\allowdisplaybreaks 
\usepackage{caption}
\usepackage[ruled,vlined,noend]{algorithm2e}

\newcommand{\mystackrel}[2]{%
	\mathrel{\vbox{\offinterlineskip\ialign{%
				\hfil##\hfil\cr
				$\scriptstyle#1$\cr
				\noalign{\kern.2ex}
				$#2$\cr
}}}}
\newboolean{useComments}
\setboolean{useComments}{FALSE}
\definecolor{dark_purple}{rgb}{0.1, 0.0, 0.4}
\definecolor{dark_green}{rgb}{0.0,0.2,0.5}
\definecolor{dark_red}{rgb}{0.85,0, 0}
\ifthenelse{\boolean{useComments}}{%
	\newcommand{\ms}[1]{\todo[inline,color=white!40,bordercolor=white]{\textcolor{teal!70!black!80}{\textbf{Manos:}\textmd{\;#1}}}} 
	\newcommand{\vh}[1]{\todo[inline,color=white!40,bordercolor=white]{\textcolor{purple!70!black!80}{\textbf{Vahid:}\textmd{\;#1}}}}  
	\newcommand{\hh}[1]{\todo[inline,color=white!40,bordercolor=white]{\textcolor{blue!70!black!80}{\textbf{Hassan:}\textmd{\;#1}}}} 
}{%
	\newcommand{\ms}[1]{}
	\newcommand{\hh}[1]{}
	\newcommand{\vh}[1]{}

}

\usepackage{bigstrut}
\usepackage{paralist}
\usepackage{tikz}
\usepackage{pgfplots}
\usepackage{comment}
\usepackage{graphicx}
\usepackage{caption}
\usepackage{ifthen}
\usepackage{paralist}
\usepackage{pgf}
\usetikzlibrary{automata,arrows,intersections,positioning,calc,shapes}
\usepackage{thm-restate}
\usepackage{etoolbox}

\renewcommand{\epsilon}{\varepsilon}
\renewcommand{\phi}{\varphi}

\newcommand{\dotteddiamond}{\kern-1pt\setbox0=\vbox{\hbox{$\diamond$}}{\ooalign{\hfil\box0\hfil\cr\hfil$\mkern-0.5mu \cdot$\hfil\crcr}}\kern-1pt}

\newcommand{\tool}{\acronym{SMC4PEP}}

\newcommand{\pb}{\acronym{pBPMN}}
\newcommand{\eb}{\acronym{eBPMN}}

\newcommand{\acronym}[1]{\ensuremath{\textnormal{#1}}}
\newcommand{\MDP}{\acronym{MDP}}

\def\middot{\textperiodcentered~} 

\usepackage{times}
\makeatletter
\RequirePackage[bookmarks,unicode,colorlinks=true]{hyperref}%
\def\@citecolor{blue}%
\def\@urlcolor{blue}%
\def\@linkcolor{blue}%

\makeatother
\title{\tool: Stochastic Model Checking of Product Engineering Processes}
\author{Hassan Hage\inst{1,2}(\Envelope) \and Emmanouil Seferis\inst{1,3} \and Vahid Hashemi\inst{2}\and Frank Mantwill\inst{1}} 
\authorrunning{H. Hage et al.}
\institute{
  Helmut-Schmidt-University, Holstenhofweg 85, 22043 Hamburg, Germany
  \email{hassan.hage@hsu-hh.de}
  \and
  AUDI AG, Auto-Union-Straße 1, 85057 Ingolstadt, Germany
  \and
  Technical University of Munich, Arcisstraße 21, 80333 Munich, Germany
}

\begin{document} 
	\maketitle 
\begin{abstract} \noindent 
	Product Engineering Processes (PEPs) are used for describing complex product developments in big enterprises such as automotive and avionics industries. The {Business Process Model Notation (BPMN)} is a widely used language to encode interactions among several participants in such PEPs. In this paper, we present \tool{} as a tool to convert graphical representations of a business process using the BPMN standard to an equivalent discrete-time stochastic control process called {Markov Decision Process (\MDP)}. To this aim, we first follow the approach described in an earlier investigation to generate a semantically equivalent business process which is more capable of handling the PEP complexity. In particular, the interaction between different levels of abstraction is realized by events rather than direct message flows. Afterwards, \tool{} converts the generated process to an \MDP{} model described by the syntax of the probabilistic model checking tool PRISM. As such, \tool{} provides a framework for automatic verification and validation of business processes in particular with respect to requirements from legal standards such as Automotive SPICE. Moreover, our experimental results confirm a faster verification routine due to smaller \MDP{} models generated from the alternative event-based BPMN models.
	\end{abstract}
	
\section{Introduction}

The ever-increasing technical challenges in products, for instance autonomous driving in automotive industries, requires \emph{Original Equipment Manufacturers (OEMs)} to restructure their \emph{Product Engineering Process (PEP)} from a mechanical-oriented to a system-oriented development to enable a rigorous verification and validation of its processes with respect to safety and non-safety requirements~\cite{Gausemeier2013SEInDerIndustriellenPraxis}. Additionally, legal authorities oblige OEMs to address consistency and traceability in their PEPs through compliance with standards such as \emph{Automotive Software Process Improvement and Capability Determination (A-SPICE)}~\cite{Aspice}. As the quality of a product is dependent on its processes's quality~\cite{Glinz2007}, consistent and qualitative processes are required for adequately addressing technical challenges, legal compliance and customer satisfaction.

A well known and most common modelling language of processes in industrial PEPs is \emph{Business Process Model and Notation (BPMN)}~\cite{OMG:objectmanagementgroup} which we refer to as \emph{pool-based BPMN (\pb)}. \pb{}s provide different users with their internal process workflows in a graphical notation and show the communication and dependency between different organization within the PEP. With the aim of facing the above mentioned challenges, the previous work in~\cite{DBLP:conf/ecsa/HageHM20} shows the need for a revision of the BPMN language which is called \emph{event-based BPMN (\eb)} in this paper. The processes, which are modelled according to the BPMN guidelines, are enriched with events and time symbols while message-flows of all processes are removed. On that way we ensure to capture time aspects like milestones of PEPs, to enable a communication between processes on different levels of abstraction by means of events, to determine the logical dependencies between processes and finally to remove process redundancies for ensuring consistency and traceability in PEPs. These argumentations on the process design motivated us to consider \eb{s} as a better design language in \tool{}. We discuss later that the \eb{} is more beneficial than its \pb{} counterpart in generating smaller \MDP{}s and hence, enabling faster verification routine. The core part of the \tool{} relies on converting \pb{}s to \eb{}s while implicitly reducing the model size which is in turn done by removing redundant processes without losing information. As a bi-product, it realizes consistency in PEPs by message passing on different levels of abstraction which is not the case if \pb{} is used as a design language. Then, \tool{} converts the generated \eb{} to an equivalent \MDP{} described in the syntax of the probabilistic model checking tool PRISM~\cite{10.1007/3-540-46002-0_5}. \tool{} ensures not only the consistency in PEPs but also allows for automated verification of generated \MDP{}s against formal description of requirements from legal standards such as A-SPICE.

\section{Related Tools}
\label{sec:3}
There exist different tools for analyzing business processes. Due to the wide industrial use of the \pb{} standard, the most common tools for analyzing business processes use this graphical representation of processes as an initial model. 

The work of Ou-Yang and Lin in~\cite{doi:10.1080/00207540701199677} provides an approach to translate \pb{}s to the Modified BPEL4WS representation and then to the Colored Petri-net XML (CPNXML) that can finally be verified by using CPN tools. This approach has restrictions in the support of split and merge conditions. The approach of Daclin et al. in~\cite{DBLP:DaclinVV15} or Mendoza Morales in~\cite{article} realize a conversion of \pb{}s to a set of Timed Automata (TA) that uses Clocked Computation Tree Logic (CCTL) for the verification. In the work of Lam in~\cite{DBLP:journals/ijseke/Lam10} \pb{}s are converted to the New Symbolic Model Verifier (NuSMV) language. Then NuSMV enables an analysis of the processes using model checking techniques and verifying properties by the Computation Tree Logic (CTL). The approaches discussed in~\cite{DBLP:DaclinVV15, DBLP:journals/ijseke/Lam10,{article},{doi:10.1080/00207540701199677}} do not consider probability distributions and non-deterministic choices of processes which are required for complex processes such as PEP.  
Duran et al.~\cite{DBLP:DuranRS18} develop the approach of Rewriting Logic to enrich \pb{}s with timing and probabilistic properties. They verify stochastic properties such as synchronization time, probability distributions by means of the Parallel Statistical Model Checking And Quantitative Analysis (PVeStA) tool. However, message passing between different processes especially on different levels of abstraction is not considered. Finally, Herbert in~\cite{DBLP:LukeH14} develop an algorithm for converting \pb{}s into MDPs, where resources like timing and probabilities are considered while message passing is performed between sub-processes. Nevertheless, the size of investigated processes is small and limited and hence, message passing between large processes in particular with different levels of abstraction is not considered. Moreover, the process model is designed with less message passing and complexity to avoid the already known state-space explosion in the generated MDP model which consequently means that this approach is not applicable on complex processes like PEPs.

\section{\tool{} Architecture and Workflow }
\label{sec:cat}
\begin{figure}
	\centering
	\includegraphics[width=0.99\linewidth]{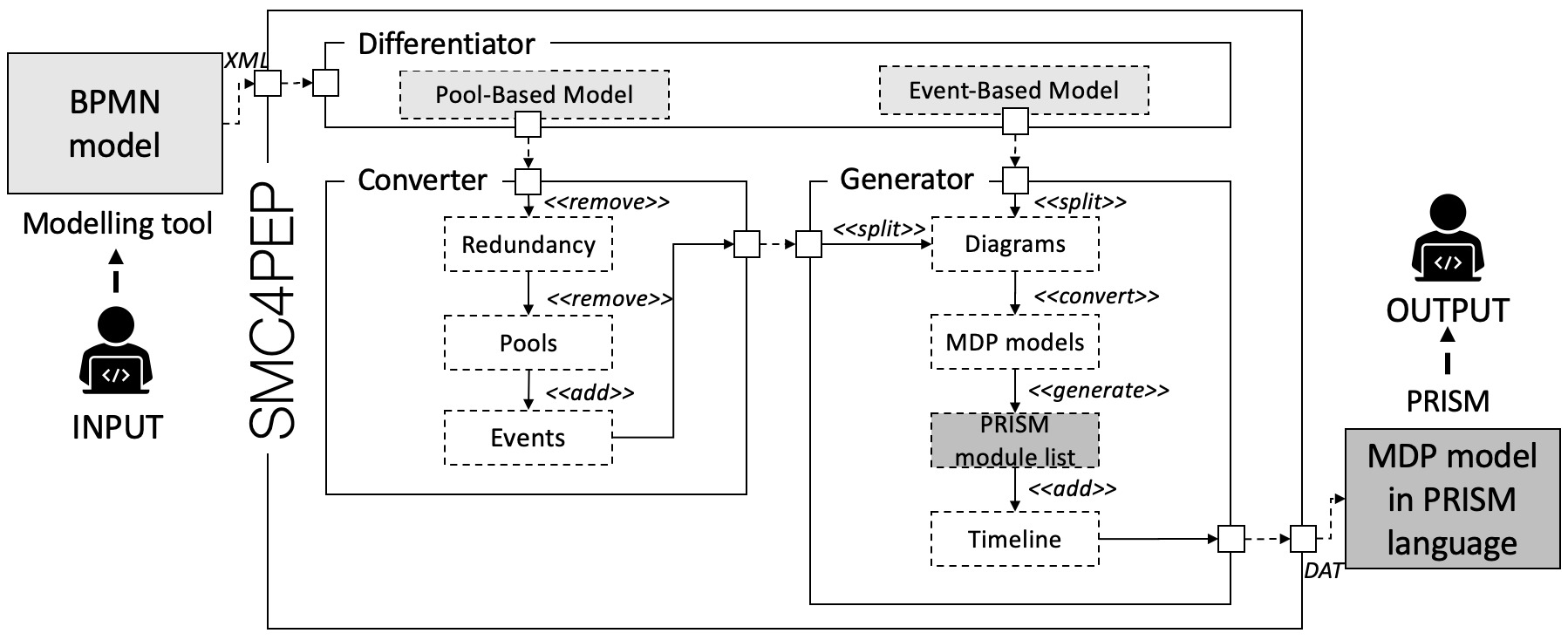}
	\caption{Architecture of the tool \tool{}.}
	\label{fig:architcturetool2}
\end{figure}

As shown in Fig.~\ref{fig:architcturetool2}, \tool{} consists of three modules, namely: (I) Differentiator, (II) Converter and (III) Generator. The Differentiator determines if the input model is a \pb{} or \eb{}. In case it is a \pb{}, the Converter converts the process model automatically to an \eb{} and moves then to the Generator. Otherwise with an \eb{} as input, \tool{} skips the Converter and moves automatically to the Generator. Finally, the Generator converts the \eb{} into an MDP described in the PRISM syntax which can directly be analyzed in PRISM. The process of generating the output PRISM model consists of three steps discussed as follows.
\paragraph{\textrm{Input.}}\tool{} requires a business process model as input with no limitation of abstraction levels. Process models can be designed either according to the guidelines in~\cite{OMG:objectmanagementgroup} or~\cite{DBLP:conf/ecsa/HageHM20} with different modelling tools such as Enterprise Architect~\cite{enterprisearchitect}. Each process model needs to be exported as an XML document for the readability of \tool{}. 

\paragraph{\textbf{\tool{}.}}The Differentiator of \tool{} receives the input document and checks the content of the BPMN model based on the syntactic and semantic differences between \eb{} and \pb{}. According to~\cite{OMG:objectmanagementgroup} message passing between processes is performed by message flows from tasks to task of the associated sub-processes, while each sub-process obtains its own boundary called pool. In the \eb{} approach message flows and pools are eliminated~\cite{DBLP:conf/ecsa/HageHM20} and each sub-process obtains its own diagram. Then the process is enriched by events to enable message passing between each process. In case of a detected \pb{}, the Differentiator triggers the Converter, otherwise the Converter will be skipped and \tool{} starts automatically the Generator. 

The Converter of \tool{} analyzes the number of identical processes within the whole process model to remove first redundant processes of \pb{} that may occur on different levels of abstraction. Redundant processes are determined when one process is equal to a second process in all elements of the model. That means in all number and content of tasks, number and content of events, number and content of gateways, role/responsible person of the process as well as number and order of sequence flows. The definition of these elements is available in~\cite{OMG:objectmanagementgroup}. When equal processes are detected, \tool{} eliminates all equal processes apart from one. Afterwards, all pools of the process models are removed and each sub-process obtains its own diagram. Finally, message flows are eliminated and replaced with events to ensure message passing and logical dependencies between the processes on different levels of abstraction. Note that message passing of the removed processes are also considered so that only one process enables a communication between different levels of abstraction. Finally, the \pb{} initial model is converted into an \eb{} and the Converter triggers the Generator.

The Generator requires an \eb{} which is provided either from the Differentiator or Converter. Then the process model is split into its number of diagrams. Afterwards, the Generator converts each diagram to an MDP taking into account message passing on different levels of abstraction by events, probability distributions and non-deterministic choices. Followed by the next step, the Generator of \tool{} generates for each MDP model a PRISM module list which are then combined to one main PRISM module list. Finally, in case of an available timeline~\cite{DBLP:conf/ecsa/HageHM20} in the process model, the PRISM module list is enriched by the values of the timeline to consider time aspects and process execution costs as rewards in the MDP model described in the PRISM syntax. 
\paragraph{\textrm{Output.}}\tool{} saves the generated MDP model described in the syntax of PRISM as a DAT document which can be uploaded into the probabilistic model checker PRISM. It is worthwhile to mention that there are quite a number of tools which are able to read the PRISM modelling language. Among others, model checkers Storm~\cite{DBLP:journals/corr/DehnertJK016}, PARAM~\cite{HahnHWZ10}, ePMC~\cite{hahn2014iscas} and Modest~\cite{hartmanns2014modest} can read our generated PRISM model for doing model checking various properties of interests. 

\section{Case Studies}
\label{sec:4}

For the evaluation of \tool{}, we converted two different use cases with \tool{}. Before, we developed an algorithm inspired by the work of~\cite{DBLP:LukeH14} to convert a \pb{} directly into an MDP. Note that this conversion is not applicable on complex processes with different levels of abstraction. Complexity means a higher number of message passing between processes, probability distributions and non-deterministic choices. \sloppy Therefore, for the evaluation we assumed that in \pb{} a communication between different levels of abstraction is possible by merging all diagrams to one main diagram, although in real processes it is not the case. This assumption is met to obtain the MDP sizes of the \pb{}. On that way MDP sizes generated through a \pb{} and \eb{} model can be compared and the effectiveness of the \eb{} can be approved. The first use case describes the process of testing an autonomous park pilot with three levels of abstraction and includes five roles where each role performs its associated task of the process.
The second use case handles a more complex process of an urgent request for a change of the vehicle construction during the PEP. In total this use case extends over four levels of abstraction and includes eleven roles. Both use cases are provided by an automotive OEM. We run all experiments on an Core i7 laptop running Windows 10.

Table~\ref{tab:table1} provides promising results generated based on \tool{}. The generated MDP model of the first use case with two levels of abstraction is for the \eb{} 33.8\% in states and 40.7\% in transitions less than for the \pb{}. Moreover, the generated MDP model in the third level of abstraction is in the \eb{} 67.78\% in states and 73.11\% in transitions less than in the \pb{}. The build time of the MDP model for the \eb{} with three levels of abstraction is higher compared to the \pb{}. Note that the MDP model is built only once which has no impact on the run-time of model checking MDPs. This is indeed the case for generating a formalism like MDP from giant BPMN models and use it several times for model checking various properties. The generated MDP models of use case two with four levels of abstraction are large compared to the first use case due to the high number of activities, probability distributions and non-deterministic choices of the processes. Nevertheless, the effectiveness of the \eb{} for complex processes is strongly confirmed by the generated MDP size of the second use case on four levels of abstraction which is far less than the MDP size of \pb{}. Finally, our generated MDP models from \eb{} have much smaller sizes compared to the approach discussed in~\cite{DBLP:LukeH14}. In particular, for the second use case we got several order of magnitudes reduction in model size which is significant for an efficient model checking routine. However, similar to~\cite{DBLP:LukeH14} we also realize the state space explosion problem which can be alleviated using bisimulation minimization techniques~\cite{hashemi2016compositional, GeblerHT12, 10.1007/978-3-319-43425-4_4}. 
\begin{table}
	\caption{Results of the analyzed processes.}
	\centering
	\begin{tabular}{c  c  c  c   c   c} 
		\hline
		BPMN & Use & Abstraction & \multicolumn{2}{c}{MDP model} & Built  \\ [0.5ex] 
		model & case & level & States & Transitions & time (s) \\ [0.5ex]
		\hline
		\pb{} & 1 & 2 & 423 & 1143 & 0.071 \\
		
		\eb{} & 1 &2 & 280 & 685 & 0.037  \\
		
		\pb{} & 1 & 3 & 5276 & 21503 & 0.170 \\
		
		\eb{} & 1 & 3 & 1700 & 5782 & 0.551  \\

		\pb{} & 2 & 4 & 93x10\textsuperscript{16} & 14x10\textsuperscript{16} &  4.263 \\
		
		\eb{} & 2 & 4 & 17x10\textsuperscript{10} & 19x10\textsuperscript{11} & 0.871 \\
		\hline
	\end{tabular}
	\label{tab:table1}
\end{table}

At the end, we take the PRISM tool for model checking some properties of interest described in the \emph{Probabilistic Reward Computation Tree Logic (PRCTL)}~\cite{DaveP11}. It is worthwhile to note that for \tool{} we provide the first use case as an \eb{} to capture time and cost aspects of the PEP by a timeline while the second use case is described first in \pb{} and then converted to \eb{}. 
\begin{table}
	\caption{Model checking of \eb{} processes.}
	\centering
	\begin{tabular}{c  c   c  c  c  c  c  c  c} 
		\hline
		Abstraction & Use & \multicolumn{2}{c}{MDP model} & \multicolumn{5}{c}{Properties} \\  
		level & Case &  States & Transitions  & $\phi_1$ & $\phi_2$ & $\phi_3$ & $\phi_4$(d) & $\phi_5$(wd)\\ [0.5ex] 
		\hline
		2 & 1 & 280 & 685 &\checkmark &\checkmark & \checkmark &78 &267.9 \\
		
		3 & 1 & 1700 & 5782 &\checkmark & \checkmark &\checkmark & 110  &346.5 \\
		
		4 & 2 & 17x10\textsuperscript{10} & 19x10\textsuperscript{11}  &\checkmark &\checkmark & \checkmark & - & - \\
		\hline
	\end{tabular}
	\label{table2}
\end{table}
Firstly, we verify some properties based on the A-SPICE guidelines~\cite{Aspice} by $\varphi_1$, $\varphi_2$ and $\varphi_3$. The properties are taken from the \emph{Generic Practice (GP)} of A-SPICE Level 2~\cite{Aspice} where each level of A-SPICE determines the quality of the processes. 
The property GP 2.1.7 of A-SPICE denoted as $\varphi_1$ which requires ensuring no deadlocks in the processes and reaching the final state of the process with the probability of 100\%. Additionally by $\varphi_2$ we denote the property GP 2.1.2 which ensures the ability of performing the process to fulfil the identified objectives similar to $\varphi_1$. Moreover, the GP 2.1.3 is denoted by $\varphi_3$ through which we ensure that our process does not deviate from its original setting according to A-SPICE.
Finally for use case one, the non-functional properties are denoted by $\varphi_4$ which delivers the minimum days (d) for performing the whole process, and by $\varphi_5$ which enables the expected cost estimation of the process obtained in accumulated working days (wd). We have to note here that $\varphi_4$ is obtained by the GUI simulator of PRISM. The results of the property verification obtained from PRISM are depicted in Table~\ref{table2}. 

\section{Conclusion}
\label{sec:5}
In this paper we presented the new tool \tool{} to enable in the first phase an automated conversion of complex process models such as PEPs that are modelled according to the BPMN standard~\cite{OMG:objectmanagementgroup} into revised process models based on the modelling approach of~\cite{DBLP:conf/ecsa/HageHM20}. This conversion paves the way for consistency and traceability of complex PEPs by removing redundant processes and enabling an exchange between different levels of process abstraction. In the second phase, \tool{} converts the new process model into an MDP to capture stochastic properties of a PEP and to enable an automated verification of the MDP using PRISM against formal descriptions of requirements. In case of designing a new PEP based on~\cite{DBLP:conf/ecsa/HageHM20}, \tool{} considers also the timeline of processes to capture time and cost aspects of a PEP that are essential for developing a new product in particular in automotive and avionics industries. Finally, we approved the effectiveness of our tool in an automotive case study where we compared \pb{s} with \eb{s} and verified some properties of interest such as legal regulations from A-SPICE.\\

\noindent\textbf{Acknowledgments.} This work is supported by the Helmut-Schmidt-University in Hamburg and by the AVAI project at AUDI AG in Ingolstadt.

\bibliographystyle{splncs04} 
\bibliography{bibtex.tex} 
\vfill

%

	
\end{document}